\documentclass[%
 reprint, superscriptaddress,
 amsmath,amssymb,
 aps,nofootinbib]{revtex4-1}
\usepackage{amsmath}
\usepackage{graphicx}
\usepackage{dcolumn}
\usepackage{bm}
\newcommand{\ket}[1]{|{#1}\rangle}
\newcommand{\bra}[1]{\langle{#1}|}
\usepackage{color}
\usepackage[colorlinks]{hyperref}
\usepackage[caption=false]{subfig}

\begin{document}

\title{Rectification of light in the quantum regime}%

\author{Jibo Dai}
 \affiliation{Centre for Quantum Technologies, National University of Singapore, 3 Science Drive 2, Singapore 117543.}
\author{Alexandre Roulet}
 \affiliation{Centre for Quantum Technologies, National University of Singapore, 3 Science Drive 2, Singapore 117543.}
\author{Huy Nguyen Le}
 \affiliation{Centre for Quantum Technologies, National University of Singapore, 3 Science Drive 2, Singapore 117543.}
\author{Valerio Scarani}
 \affiliation{Centre for Quantum Technologies, National University of Singapore, 3 Science Drive 2, Singapore 117543.}
 \affiliation{Department of Physics, National University of Singapore, 2 Science Drive 3, Singapore 117542.}

\date{\today}

\begin{abstract}
One of the missing elements for realising an integrated optical circuit is a rectifying device playing the role of an \emph{optical diode}. A proposal based on a pair of two-level atoms strongly coupled to a one-dimensional waveguide showed a promising behavior based on a semi-classical study [Fratini et al., Phys. Rev. Lett. \textbf{113}, 243601 (2014)]. Our study in the full quantum regime shows that, in such a device, rectification is a purely multi-photon effect. For an input field in a coherent state, rectification reaches up to $70\%$ for the range of power in which one of the two atoms is excited, but not both.
\end{abstract}

\maketitle


On the road towards a quantum network \cite{Kimble} built with integrated optoelectronic components, several devices have been proposed and studied \cite{Hoi+5:11, Sala+Blaauboer:15}. Two-level emitters (atoms) coupled strongly to one-dimensional (1D) waveguide have recently emerged as a prominent candidate in building this integrated network. However, one of the key elements that is still missing for such an integrated circuit is a rectifying device a.k.a. \emph{optical diode}, which could be used for instance as an on-chip optical isolator \cite{Roy:10, Roy:13, Jala+12:13}. Obviously, one wants a \textit{passive} diode: if the state of the rectifier is correlated to the input direction of light, rectification is trivial, but coherence in the output field is lost. The first passive light diode to be proposed relied on the light fields impinging from opposite directions having been prepared in orthogonal polarisation \cite{Shen+2:11}. Ultimately one would also want \textit{state-independent} rectification.

\begin{figure}
\subfloat{
\includegraphics[width=0.47\textwidth]{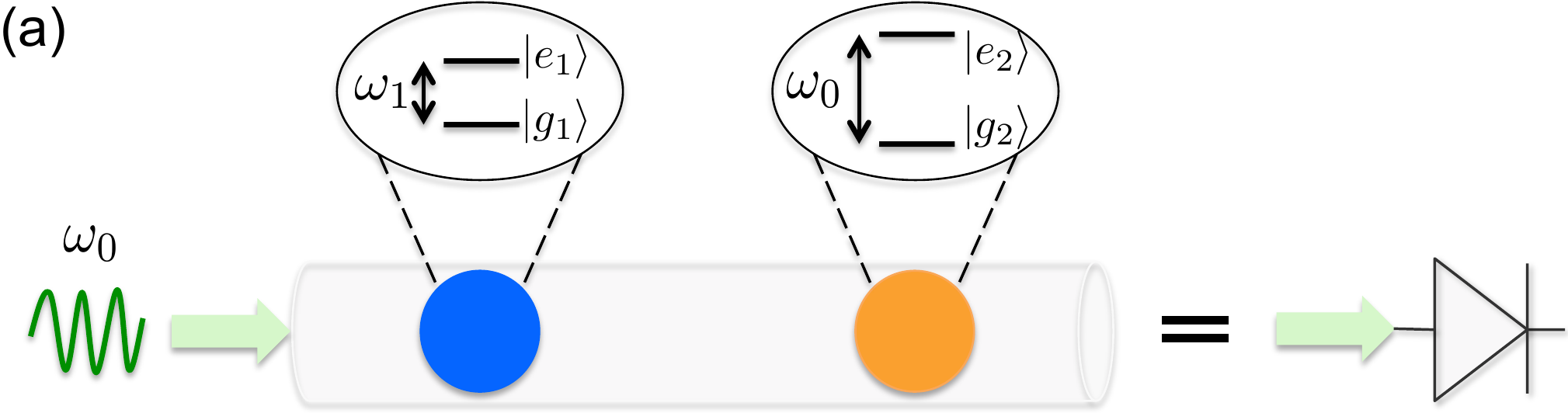}\label{fig:DiodeG}
}\quad
\subfloat{
\includegraphics[width=0.47\textwidth]{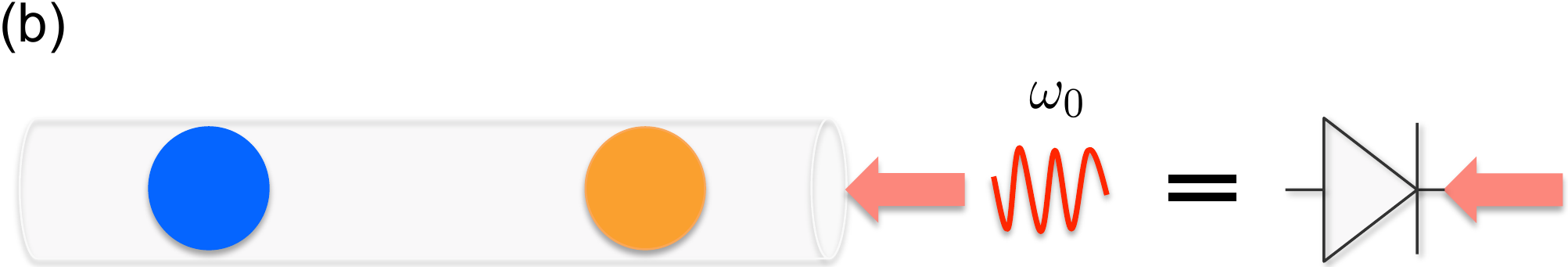}\label{fig:DiodeR}
}
\caption{\label{fig:Diode}(color online).\quad A light pulse impinging on an optical diode formed by a pair of two-level atoms strongly coupled to a 1D waveguide. The atom on the right is resonant with the pulse. (a) When coming from the left, the pulse is transmitted through the diode. (b) Conversely, a pulse entering from the right is reflected, preventing contamination of the circuit.}
\end{figure}

A possible candidate for a passive and state-independent optical diode, was identified recently \cite{Fratini+7:14}: it is the analog of a Fabry-Perot interferometer built with two atoms with different transition frequencies, coupled to a 1D waveguide (Fig.~\ref{fig:Diode}). A semi-classical analysis predicted a rectification factor as high as $92 \%$ in some power range. In this paper, we present a full quantum mechanical analysis of this setup, both for coherent input pulse and single photon input pulse. We find that the setup under study cannot achieve rectification for single-photon states, contrary to the hope stated in the initial proposal. When the incident light is a coherent pulse, the rectification factor can reach up to $\sim 70\%$ in a range of power, whereas no rectification is predicted for low and high power. By studying the dynamics of excitation of the two atoms, we obtain a clear physical picture for the origin of the rectification.

While completing this paper, we became aware that two other groups reported the same study, both using the formalism of master equations \cite{Fratini+Ghobadi:15,Mascarenhas+3:15}. Our work, which is rather based on the Heisenberg equations of motion, is completely independent of theirs, but we have formatted our plots as to facilitate comparisons and cross-validation with Ref.~\cite{Fratini+Ghobadi:15}.

\textit{Model.---} We consider the system illustrated in Fig.~\ref{fig:Diode}, consisting of a pair of atoms strongly coupled to a 1D waveguide and separated by a distance $d=|x_2-x_1|$, where $x_j$ is the position of the $j$-th atom. Both atoms are modelled as two-level systems, denoted as usual by $|\text{g}_j\rangle$ for the ground and $|\text{e}_j\rangle$ for the excited state. The first atom has resonance frequency $\omega_1$, while the second atom is on resonance with the incident light pulse centred around the frequency $\omega_0$. We assume that $\omega_1$ and $\omega_0$ are much larger than the cutoff frequency of the waveguide, such that the longitudinal wave number $k$ of the photon mode obeys the linearized dispersion relation $\omega=v_\text{g}|k|$, where $v_\text{g}$ is the group velocity of the photon in the waveguide~\cite{Snyder+Love}.

This system has been studied with various theoretical tools: post-scattering descriptions based on a real-space formalism \cite{Shen+Fan:07-1, Shen+Fan:07-2, Zheng+2:10}, input-output theory \cite{Fan+2:10}, standard scattering theory and generalized master equations \cite{Pletyukhov+Gritsev, Shi+2}. Here, our approach is to use Heisenberg equations of motion, where the dipole Hamiltonian describing the interaction between the atoms and propagating photons, under rotating wave approximation, is given by~\cite{Domokos+2:02,Roulet:14}
\begin{align}\nonumber
\hat{H}_{\text{dip}}=&-\text{i}\hbar\sum_{j=1}^2\int_{0}^\infty\!\text{d} \omega\, g^{(j)}_\omega\\
&\times\left[\hat{\sigma}^{(j)}_+\left(\hat{a}_\omega \text{e}^{\text{i}\omega x_j/v_\text{g}}+\hat{b}_\omega \text{e}^{-\text{i}\omega x_j/v_\text{g}}\right)-\text{H}.\text{c}.\right]\ ,
\end{align}
where $g_\omega^{(j)}$ is the coupling constant of the $j$th atom with raising ladder operator $\hat{\sigma}_+^{(j)}=\ket{\text{e}_j}\bra{\text{g}_j}$ and $\hat{a}_\omega$ ($\hat{b}_\omega$) is the annihilation operator of the forward- (backward-) propagating photon mode at frequency $\omega$. For simplicity, we will assume identical coupling constants in the Weisskopf-Wigner approximation \cite{Scully+Zubairy}, $g_\omega^{(1)}=g_\omega^{(2)}=g$, yielding a decay rate to the waveguide of $\gamma=2\pi g^2$ for each atom (the case of different values of $\gamma$ for the two atoms is presented in Appendix A). Moreover, based upon the high coupling efficiencies achieved by artificial atoms such as quantum dots \cite{Claudon+8:10, Arcari+10:14} and superconducting qubits in 1D geometry \cite{Hoi+5:11, Loo+5:13}, we will focus on the ideal situation in the absence of loss. Significant progress is also being made towards strong coupling between light and atoms \cite{Goban+6:15} as well as nitrogen vacancy centres coupled to 1D surface plasmon \cite{Huck+3:11}. Any relevant physical quantity is then readily obtained by deriving the Heisenberg equation of the corresponding observable and solving a closed set of first-order differential equations (see Appendix A for further details on the derivations). 

We study the behaviour of the alleged optical diode when shined with either a single-photon pulse or a coherent pulse of frequency bandwidth $\Omega\ll\gamma$. For the latter we use the multimode definition of a coherent state delivering a constant mean photon flux $|\alpha|^2$ with Poissonian statistics~\cite{Loudon:90} (see Appendix A for more details). We can then obtain the fraction of light transmitted (reflected) by the diode in the steady-state regime when the light is injected from the left (right) as
\begin{equation}
\bar{N}_{b}=\frac{1}{|\alpha|^2}\int_0^\infty\text{d}\omega\langle\hat{b}_\omega^\dagger\hat{b}_\omega\rangle_\text{ss}.
\end{equation}
Importantly, the atoms are initially in the ground state independently of the incoming light pulse in order to ensure the passive attribute of the rectifying device. Similar to previous studies~\cite{Lepri+Casati:11, Fratini+7:14}, the figure of merit characterising the diode efficiency is then defined as
\begin{equation}
\mathcal{L}=\frac{|T_{\rightarrow}-T_{\leftarrow}|}{T_{\rightarrow}+T_{\leftarrow}}T_{\rightarrow},
\end{equation}
where $T_{\rightarrow}=1-\bar{N}_{b}$ is the transmittance for the case when light is incident from left (Fig.~\ref{fig:DiodeG}) while $T_{\leftarrow}=\bar{N}_{b}$ is for the reversed situation (Fig.~\ref{fig:DiodeR}). A large efficiency $\mathcal{L}$ ensures both strong directionality and significant transmission of light when coming from the left.

\begin{figure}
\subfloat{
\includegraphics[width=0.22\textwidth]{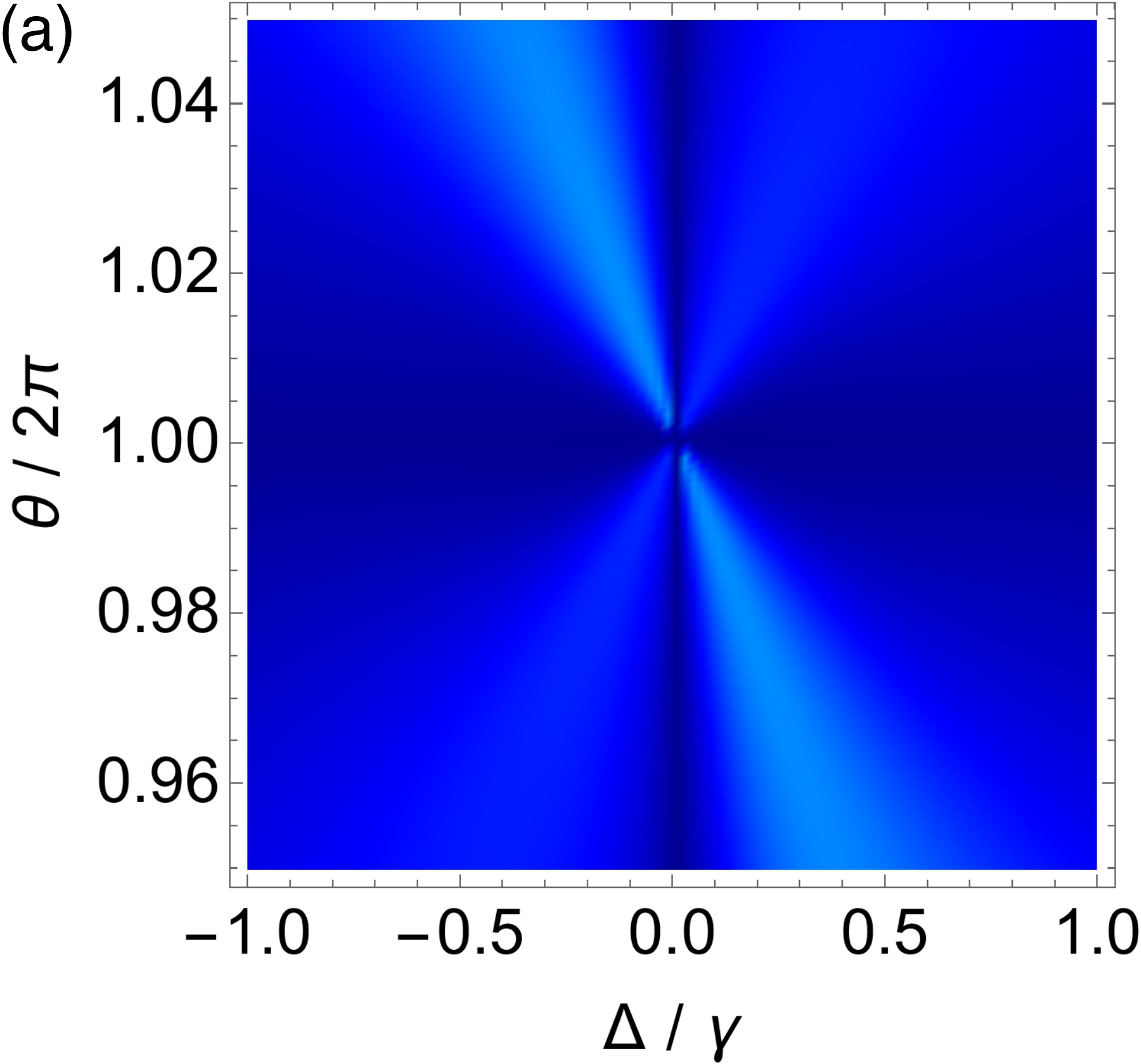}\label{fig:Density1}
}\quad
\subfloat{
\includegraphics[width=0.22\textwidth]{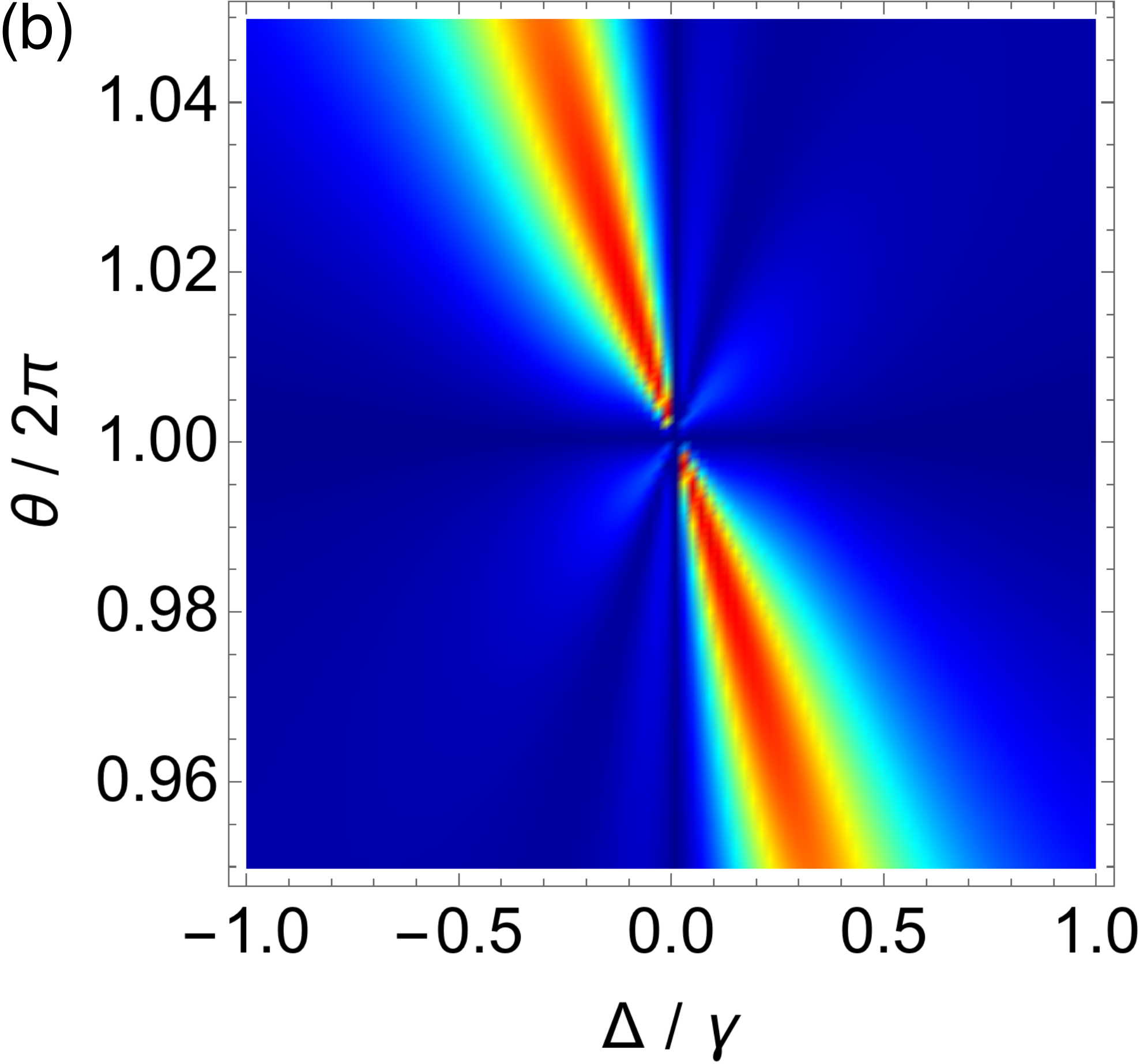}\label{fig:Density2}
}\\
\subfloat{
\includegraphics[width=0.22\textwidth]{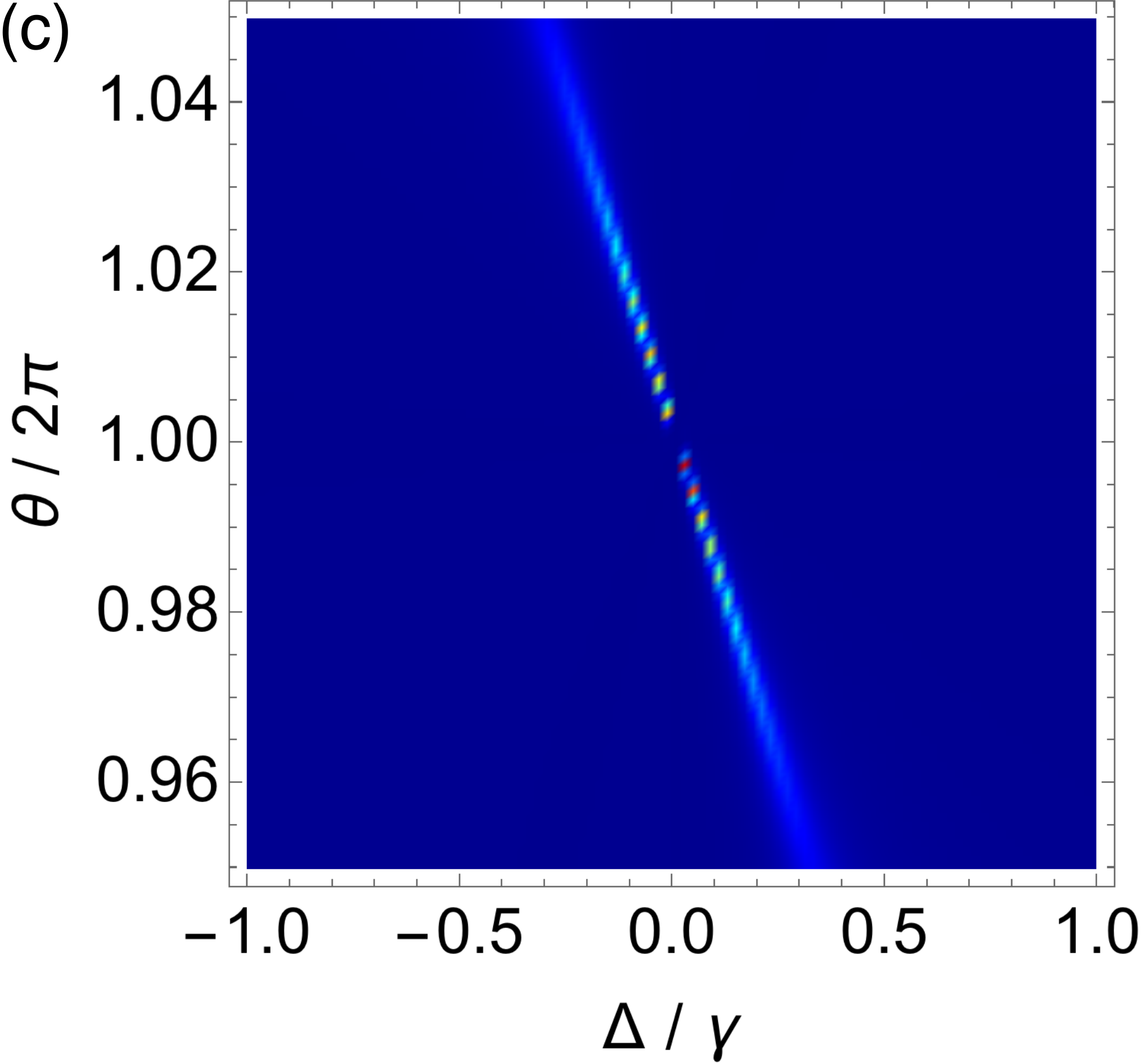}\label{fig:Density3}
}\quad
\subfloat{
\includegraphics[width=0.22\textwidth]{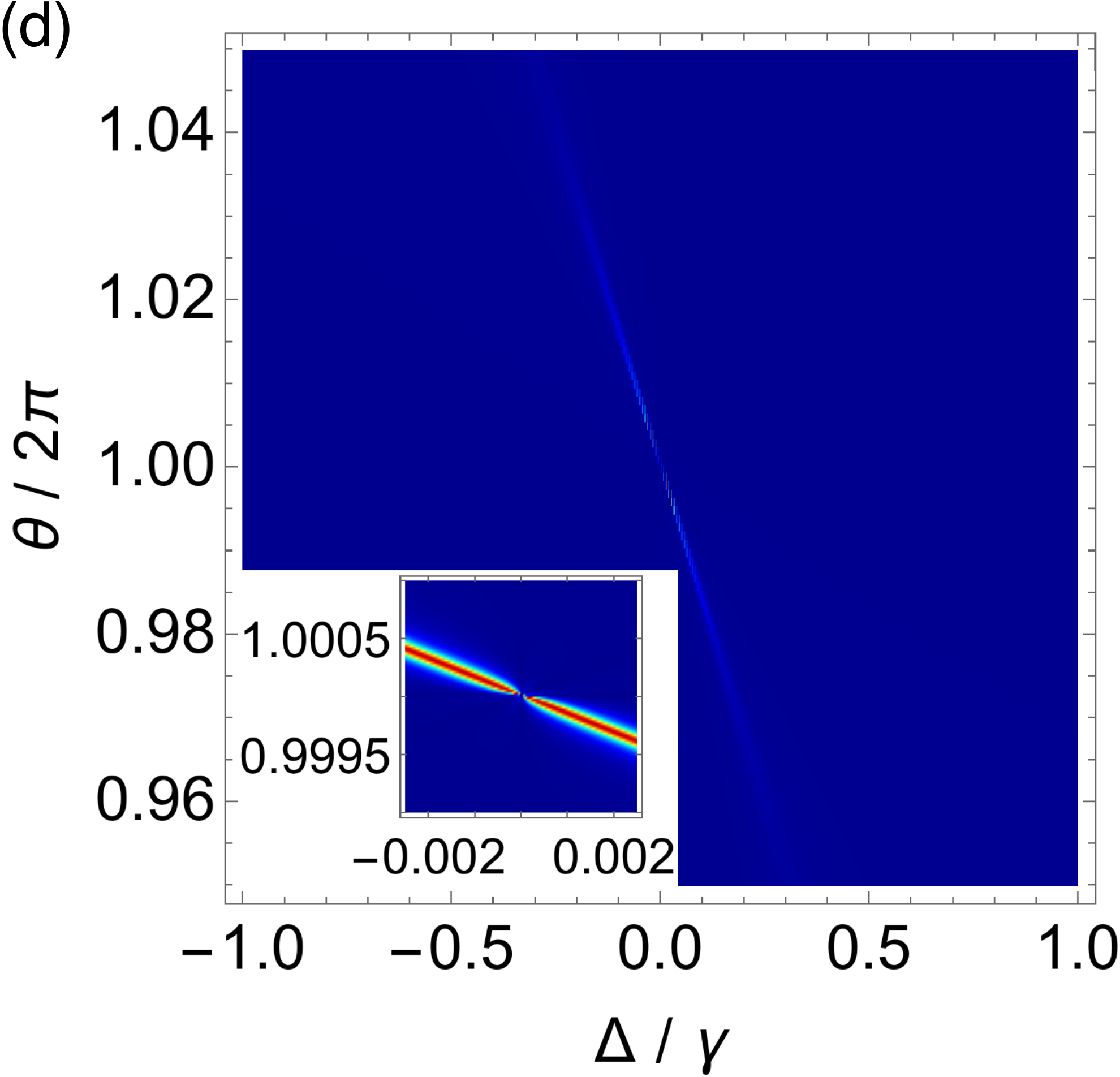}\label{fig:Density4}
}\vskip 0.1cm
\subfloat{
\includegraphics[width=0.15\textwidth]{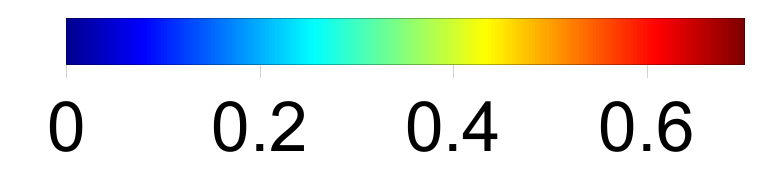}\label{fig:DensityScale}
}
\caption{\label{fig:DensityPlot}(color online).\quad The diode efficiency $\mathcal{L}$ for different pump powers as a function of the detuning of the atom on the left, $\Delta=\omega_0-\omega_1$, and the phase $\theta=2\omega_0 d/v_\text{g}$. (a) $|\alpha|^2/\gamma=1$. (b) $|\alpha|^2/\gamma=10^{-1}$. (c) $|\alpha|^2/\gamma=10^{-3}$. (d) $|\alpha|^2/\gamma=10^{-4}$.}
\end{figure}

\begin{figure}
\includegraphics[width=0.45\textwidth]{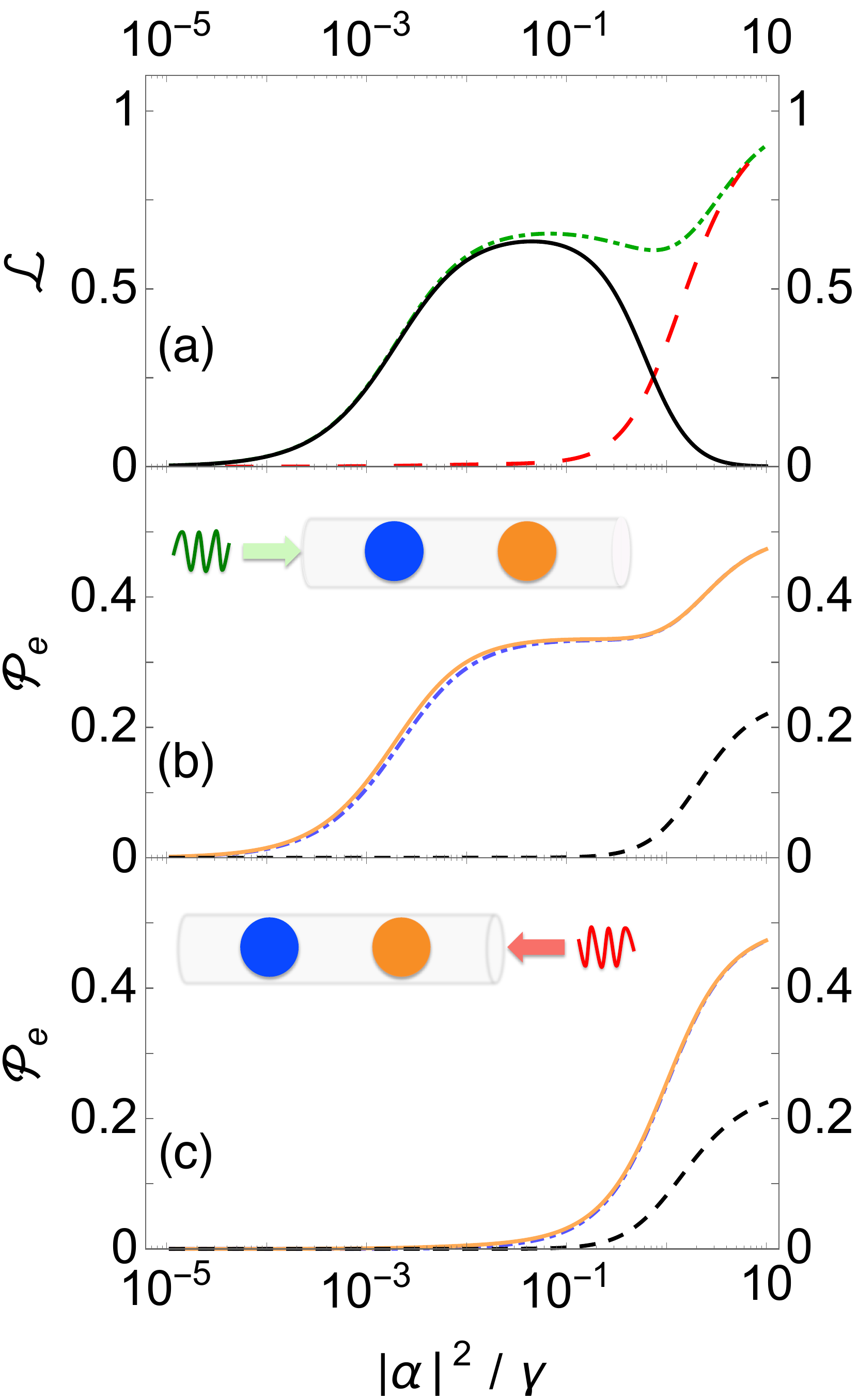}
\caption{\label{fig:FigPower}(color online).\quad The diode efficiency as a function of the mean number of photons per atomic lifetime for fixed detuning $\Delta/\gamma=0.12$ and phase $\theta=2\pi\times 0.982$ (same detuning parameter as in Fig.~4 of Ref.~\cite{Fratini+Ghobadi:15}). (a) The diode efficiency $\mathcal{L}$ in black is plotted together with $T_{\rightarrow}$ in dashed dotted green and $T_{\leftarrow}$ in dashed red. (b) Light is injected from the left. In orange (dashed dotted blue) is the probability of excitation of the atom on the right (left), and in dashed black the probability of having both atoms excited at the same time. (c) Same as (b) but for light injected from the right. This indicates that rectification happens for light that comes when only one atom is excited. At low power, no atom is excited and the diode is reflecting from both directions; at high power, both atoms are excited and the diode is transparent from both directions.}
\end{figure}

\textit{Results.---} For \textit{single-photon pulses}, the rectification is found to be negligible (see Appendix B for details). Indeed, denoting by $\Delta_j$ the detuning of the $j$-th atom with the pulse, the reflectivity of the diode is found to be
\begin{align}
R=\frac{\left[(\Delta_1+\Delta_2)\cos \tfrac{\theta}{2}+2\sin \tfrac{\theta}{2}\right]^2+\left[(\Delta_2-\Delta_1)\sin\tfrac{\theta}{2}\right]^2}{\left[\Delta_1\Delta_2-1+\cos \theta\right]^2+\left[\Delta_1+\Delta_2+\sin \theta\right]^2}, \label{T_formula}
\end{align}
which is symmetric upon exchanging $\Delta_1$ and $\Delta_2$. Hence, $R$ is independent of the input direction of the photon: such a set-up cannot rectify light propagation in the single-photon regime. Notice that these calculations agree with the fully symmetric transmission coefficient obtained in Ref.~\cite{Zheng+Baranger} using post-scattering descriptions based on a real-space formalism.

Fortunately, the diode does achieve rectification when the inputs are \textit{coherent states}. In Fig.~\ref{fig:DensityPlot}, we plot the diode efficiency $\mathcal{L}$ for coherent states of different mean photon flux $|\alpha|^2$ as a function of the detuning $\Delta=\omega_0-\omega_1$ and the phase parameter $\theta=2\omega_0 d/v_\text{g}$. As a first observation, our results agree qualitatively with the semi-classical calculation based on a Fabry-Perot interferometer \cite{Fratini+7:14} where considerable directionality was obtained for a pumping power of $|\alpha|^2 / \gamma=10^{-1}$. However, the efficiency does not exceed $70 \%$ (see Fig.~\ref{fig:Density2}), in contrast with the semi-classical prediction that exceeded $92 \%$. Moreover, we notice another feature that was not discussed previously: when the power is further decreased (Fig.~\ref{fig:Density3} and Fig.~\ref{fig:Density4}), the efficiency is found to drop drastically over most of the parameter space. One can still find small regions of high efficiency, but it would require extreme fine-tuning to select those working points when $|\alpha|^2 / \gamma\lesssim 10^{-4}$.

In order to investigate how the efficiency varies with the power of the input coherent state, we study the diode response for fixed values of the detuning $\Delta$ and the phase $\theta$. The results are shown in Fig.~\ref{fig:FigPower}a. Firstly, one notice that as the power increases beyond approximately $|\alpha|^2 / \gamma\approx 10^{-1}$, the diode efficiency starts to drop. This is easily understood as being a consequence of the fact that the atoms can absorb at most one photon. Thus when a significant mean number of photon per atomic lifetime enters the device, both atoms are highly saturated and most of the light is transmitted regardless of which side it comes from. On the other hand, the efficiency also goes to zero as the input power tends towards zero, in agreement with the single-photon result.

\textit{Atom excitation as the origin of rectification.---} 

Both the single-photon result and the absence of rectification for very low power indicate clearly that rectification is due to multi-photon components in the light field~\footnote{Note that since the global phase of the coherent state is irrelevant, we can reason in terms of a Poissonian mixture of photon number states \cite{Molmer:97}.}. While this is clearly established, looking at the horizontal axis of Fig.~\ref{fig:FigPower}a, one may wonder what multi-photon effects one can expect for a coherent state of power as small as $|\alpha|^2 /\gamma \approx 10^{-3}$. In other words, the qualitative behavior being admitted, one might have expected to find the effective rectification range at higher powers. We are going to provide evidence that rectification becomes effective when one atom is excited but not both. Obviously this mechanism requires at least two photons in the field: one to excite one atom, the other(s) to be transmitted or reflected.

Let us then study the excitation probabilities of the atoms in the steady state. Specifically, we monitor the probability of finding the first (second) atom excited $\mathcal{P}^e_1$ ($\mathcal{P}^e_2$), or of finding both atoms excited simultaneously $\mathcal{P}^e_{12}$. We calculate these probabilities for the case where light is incident from the left (Fig.~\ref{fig:FigPower}b), as well as from the right (Fig.~\ref{fig:FigPower}c). We first observe that $\mathcal{P}^e_{12}$ is found to be independent of where the light comes from, and starts to increase significantly at input power around $|\alpha|^2 / \gamma\approx 1$. This coincides with the drop in rectification efficiency at the high power end, thus matching the explanation given above: the diode becomes transparent from both directions when both atoms are significantly excited.

On the other hand, while $\mathcal{P}^e_1$ and $\mathcal{P}^e_2$ are found to be approximately equal (compare the two curves in each panel), they vary significantly depending on the direction of the incoming pulse (compare the two panels). When the light is coming from the right, it encounters first the resonant atom which acts as an almost perfect mirror when $|\alpha|^2 / \gamma\ll 1$. This can be seen from the reflection coefficient of a coherent state on a single atom
\begin{equation}
R=\frac{1}{1+(\Delta/\gamma)^2+2|\alpha|^2/\gamma}. \label{Rfor1Atom}
\end{equation}
which is in agreement in the limit of low power with the single-photon result \cite{Shen+Fan:05}. Hence we get $T_{\leftarrow}\approx 0$ and the off-resonant atom does not play any role. Only when the power becomes non-negligible $|\alpha|^2 / \gamma\approx 1$ will the right atom start saturating. However at this level of power both atoms start to be significantly excited and no directionality is expected as explained previously.

When the light is coming from the left, the behavior is richer. Both $\mathcal{P}^e_1$ and $\mathcal{P}^e_2$ start increasing at a very low power of $|\alpha|^2 / \gamma\approx 10^{-4}$, reaching a plateau until a power of $|\alpha|^2 / \gamma\approx 1$. This can understood as follow: thanks to the non-zero detuning of the first atom, the light is able to enter the cavity and is stored inside for some time. The resonant atom then saturates at a lower power than when light is incident from the right. We identify this as the key mechanism leading to rectification of light in the range of power corresponding to this plateau. At lower power, the light is not stored for long enough time, so the device effectively sees one photon at a time, and the single photon result with no directionality is recovered.

\textit{Conclusions.---}
We presented a full quantum mechanical analysis of a proposed optical diode consisting of two different atoms coupled to a 1D waveguide using Heisenberg equations. We found that the diode fails to work in the single-photon regime but performs rectification for input coherent states in an optimal range of power. The working mechanism of the diode is revealed by studying the excitation probabilities of the two atoms. This detailed understanding may inspire improved designs of passive state-independent optical diodes.

\begin{acknowledgements}
We acknowledge clarifying discussions with Marcelo Fran\c{c}a Santos.
This research is supported by the National Research Foundation (partly through its Competitive Research Programme, Award No. NRF-CRP12-2013-03) and the Ministry of Education, Singapore.
\end{acknowledgements}

\begin{widetext}
\appendix
\numberwithin{equation}{section}
\section{Coherent pulse}
For coherent pulse, we work in the Heisenberg picture. We derive the equations of motion in the more general case where $\gamma_1=2\pi\left(g^{(1)}\right)^2$ and $\gamma_2=2\pi\left(g^{(2)}\right)^2$ may not be the same. To describe the evolution of the system during the scattering event, it is convenient to work in the interaction picture with respect to the free Hamiltonian $\hat{H}_0=\hat{H}_\text{atom}+\hat{H}_\text{field}$, where $\displaystyle{\hat{H}_\text{atom}=\sum_{j=1}^2\hbar\omega_j|\text{e}_j\rangle\langle \text{e}_j|}$, and $\displaystyle{\hat{H}_\text{field}=\int_0^\infty\text{d}\omega \hbar\omega \left(\hat{a}_\omega^\dagger\hat{a}_\omega+\hat{b}_\omega^\dagger\hat{b}_\omega\right)}$. The Hamiltonian in the interaction picture then takes the form of Eq. (1) in the main text. Making the Weisskopf-Wigner approximation \cite{Scully+Zubairy}, the field operators evolve as
\begin{equation}
\hat{a}_\omega(t)=\hat{a}_\omega(0)+g_\omega^{(1)}\int_0^t\text{d}t'\hat{\sigma}_-^{(1)}(t')\text{e}^{\text{i}(\omega-\omega_1)t'}+g_\omega^{(2)}\int_0^t\text{d}t'\hat{\sigma}_-^{(2)}(t')\text{e}^{\text{i}[(\omega-\omega_2)t'-\omega d/v_\text{g}]},
\end{equation}
and
\begin{equation}
\hat{b}_\omega(t)=\hat{b}_\omega(0)+g_\omega^{(1)}\int_0^t\text{d}t'\hat{\sigma}_-^{(1)}(t')\text{e}^{\text{i}(\omega-\omega_1)t'}+g_\omega^{(2)}\int_0^t\text{d}t'\hat{\sigma}_-^{(2)}(t')\text{e}^{\text{i}[(\omega-\omega_2)t'+\omega d/v_\text{g}]}.
\end{equation}
Substituting the field operators into the equations for the atomic operators, one obtains a closed set of equations for the atomic operators in the Markovian regime. In this regime, we neglected the time delay $d/v_\text{g}$ induced by the distance between the two atoms. This is justified as under realistic experimental set-up, $d/v_\text{g}$ is many orders of magnitude smaller than the timescale set by the inverse of the interaction strength $1/\gamma$, which governs the timescale at which the system evolves \cite{Hoi+5:11, Goban+6:15}. Hence, the operator equations are given by, with $\tau=d/v_\text{g}$, $\Delta_{12}=\Delta_1-\Delta_2$, $\theta_1=\omega_1\tau$ and $\theta_2=\omega_2\tau$
\begin{equation}
\frac{\text{d}}{\text{d}t}\hat{\sigma}_-^{(1)}=-\gamma_1\hat{\sigma}_-^{(1)}+\sqrt{\gamma_1\gamma_2}\hat{\sigma}_z^{(1)}\hat{\sigma}_-^{(2)}\text{e}^{-\text{i}\Delta_{12} t}\text{e}^{\text{i}\theta_2}+\sqrt{\gamma_1}\hat{\sigma}_z^{(1)}\left(\hat{a}_t+\hat{b}_t\right).
\end{equation}
\begin{equation}
\frac{\text{d}}{\text{d}t}\hat{\sigma}_z^{(1)}=-2\gamma_1\left(\hat{\sigma}_z^{(1)}+\mathbf{1}\right)-2\sqrt{\gamma_1\gamma_2}\left(\hat{\sigma}_+^{(1)}\hat{\sigma}_-^{(2)}\text{e}^{-\text{i}\Delta_{12} t}\text{e}^{\text{i}\theta_2}+\text{H.c}.\right)-2\sqrt{\gamma_1}\left(\hat{\sigma}_+^{(1)}(\hat{a}_t+\hat{b}_t)+\text{H.c}.\right).
\end{equation}
\begin{equation}
\frac{\text{d}}{\text{d}t}\hat{\sigma}_-^{(2)}=-\gamma_2\hat{\sigma}_-^{(2)}+\sqrt{\gamma_1\gamma_2}\hat{\sigma}_z^{(2)}\hat{\sigma}_-^{(1)}\text{e}^{\text{i}\Delta_{12} t}\text{e}^{\text{i}\theta_1}+\sqrt{\gamma_2}\hat{\sigma}_z^{(2)}\left(\hat{a}_{t-\tau} \text{e}^{\text{i}\theta_1}+\hat{b}_{t+\tau} \text{e}^{-\text{i}\theta_1}\right)\text{e}^{\text{i}\Delta_{12} t}.
\end{equation}
\begin{equation}
\frac{\text{d}}{\text{d}t}\hat{\sigma}_z^{(2)}=-2\gamma_2\left(\hat{\sigma}_z^{(2)}+\mathbf{1}\right)-2\sqrt{\gamma_1\gamma_2}\left(\hat{\sigma}_+^{(2)}\hat{\sigma}_-^{(1)}\text{e}^{\text{i}\Delta_{12} t}\text{e}^{\text{i}\theta_1}+\text{H.c}.\right)-2\sqrt{\gamma_2}\left(\hat{\sigma}_+^{(2)}\left(\hat{a}_{t-\tau} \text{e}^{\text{i}\theta_1}+\hat{b}_{t+\tau} \text{e}^{-\text{i}\theta_1}\right)\text{e}^{\text{i}\Delta_{12} t}+\text{H.c}.\right).
\end{equation}
\begin{align}\nonumber
\frac{\text{d}}{\text{d}t}\left(\hat{\sigma}_z^{(1)}\hat{\sigma}_-^{(2)}\right)=(-2\gamma_1-\gamma_2)\hat{\sigma}_z^{(1)}\hat{\sigma}_-^{(2)}-2\gamma_1\hat{\sigma}_-^{(2)}-\sqrt{\gamma_1\gamma_2}\hat{\sigma}_-^{(1)}\text{e}^{\text{i}\Delta_{12} t}\text{e}^{-\text{i}\theta_2}-\sqrt{\gamma_1\gamma_2} \hat{\sigma}_-^{(1)}\hat{\sigma}_z^{(2)}\text{e}^{\text{i}\Delta_{12} t}\left(\text{e}^{-\text{i}\theta_2}+\text{e}^{\text{i}\theta_1}\right)\\
-2\sqrt{\gamma_1}\left(\hat{\sigma}_+^{(1)}\hat{\sigma}_-^{(2)}\left(\hat{a}_t+\hat{b}_t\right)+\left(\hat{a}^\dagger_t+\hat{b}^\dagger_t\right)\hat{\sigma}_-^{(1)}\hat{\sigma}_-^{(2)}\right)+\sqrt{\gamma_2}\hat{\sigma}_z^{(1)}\hat{\sigma}_z^{(2)}\text{e}^{\text{i}\Delta_{12} t}\left(\hat{a}_{t-\tau} \text{e}^{\text{i}\theta_1}+\hat{b}_{t+\tau} \text{e}^{-\text{i}\theta_1}\right).
\end{align}
\begin{align}\nonumber
\frac{\text{d}}{\text{d}t}\left(\hat{\sigma}_-^{(1)}\hat{\sigma}_z^{(2)}\right)=(-2\gamma_2-\gamma_1)\hat{\sigma}_-^{(1)}\hat{\sigma}_z^{(2)}-2\gamma_2\hat{\sigma}_-^{(1)}-\sqrt{\gamma_1\gamma_2}\hat{\sigma}_-^{(2)}\text{e}^{-\text{i}\Delta_{12} t}\text{e}^{-\text{i}\theta_1}-\sqrt{\gamma_1\gamma_2} \hat{\sigma}_z^{(1)}\hat{\sigma}_-^{(2)}\text{e}^{-\text{i}\Delta_{12} t}\left(\text{e}^{-\text{i}\theta_1}+\text{e}^{\text{i}\theta_2}\right)\\
-2\sqrt{\gamma_2}\left(\hat{\sigma}_-^{(1)}\hat{\sigma}_+^{(2)}\left(\hat{a}_{t-\tau} \text{e}^{\text{i}\theta_1}+\hat{b}_{t+\tau} \text{e}^{-\text{i}\theta_1}\right)\text{e}^{\text{i}\Delta_{12} t}+\left(\hat{a}^\dagger_{t-\tau} \text{e}^{-\text{i}\theta_1}+\hat{b}^\dagger_{t+\tau} \text{e}^{\text{i}\theta_1}\right)\text{e}^{-\text{i}\Delta_{12} t}\hat{\sigma}_-^{(1)}\hat{\sigma}_-^{(2)}\right)+\sqrt{\gamma_1}\hat{\sigma}_z^{(1)}\hat{\sigma}_z^{(2)}\left(\hat{a}_t +\hat{b}_t\right).
\end{align}
\begin{align}\nonumber
\frac{\text{d}}{\text{d}t}\left(\hat{\sigma}_+^{(1)}\hat{\sigma}_-^{(2)}\right)=&-(\gamma_1+\gamma_2)\hat{\sigma}_+^{(1)}\hat{\sigma}_-^{(2)}+\frac{1}{2}\sqrt{\gamma_1\gamma_2} \text{e}^{\text{i}\Delta_{12} t}\left(\hat{\sigma}_z^{(1)}\text{e}^{-\text{i}\theta_2}+\hat{\sigma}_z^{(2)}\text{e}^{\text{i}\theta_1}\right)+\frac{1}{2}\sqrt{\gamma_1\gamma_2}\hat{\sigma}_z^{(1)}\hat{\sigma}_z^{(2)}\text{e}^{\text{i}\Delta_{12} t}\left(\text{e}^{\text{i}\theta_1}+\text{e}^{-\text{i}\theta_2}\right)\\
&+\sqrt{\gamma_1}\left(\hat{a}^\dagger_t+\hat{b}^\dagger_t\right)\hat{\sigma}_z^{(1)}\hat{\sigma}_-^{(2)}+\sqrt{\gamma_2}\hat{\sigma}_+^{(1)}\hat{\sigma}_z^{(2)}\text{e}^{\text{i}\Delta_{12} t}\left(\hat{a}_{t-\tau} \text{e}^{\text{i}\theta_1}+\hat{b}_{t+\tau} \text{e}^{-\text{i}\theta_1}\right).
\end{align}
\begin{align}
\frac{\text{d}}{\text{d}t}\left(\hat{\sigma}_-^{(1)}\hat{\sigma}_-^{(2)}\right)=-(\gamma_1+\gamma_2)\hat{\sigma}_-^{(1)}\hat{\sigma}_-^{(2)}+\sqrt{\gamma_1}\hat{\sigma}_z^{(1)}\hat{\sigma}_-^{(2)}\left(\hat{a}_t+\hat{b}_t\right)+\sqrt{\gamma_2}\hat{\sigma}_-^{(1)}\hat{\sigma}_z^{(2)}\text{e}^{\text{i}\Delta_{12} t}\left(\hat{a}_{t-\tau} \text{e}^{\text{i}\theta_1}+\hat{b}_{t+\tau} \text{e}^{-\text{i}\theta_1}\right).
\end{align}
\begin{align}\nonumber
\frac{\text{d}}{\text{d}t}\left(\hat{\sigma}_z^{(1)}\hat{\sigma}_z^{(2)}\right)=-2(\gamma_1+\gamma_2)\hat{\sigma}_z^{(1)}\hat{\sigma}_z^{(2)}-2\gamma_2\hat{\sigma}_z^{(1)}-2\gamma_1\hat{\sigma}_z^{(2)}+2\sqrt{\gamma_1\gamma_2}\left(\hat{\sigma}_+^{(1)}\hat{\sigma}_-^{(2)}\text{e}^{-\text{i}\Delta_{12} t}\left(\text{e}^{-\text{i}\theta_1}+\text{e}^{\text{i}\theta_2}\right)+\text{H.c.}\right)\\
-2\sqrt{\gamma_1}\left(\hat{\sigma}_+^{(1)}\hat{\sigma}_z^{(2)}\left(\hat{a}_t+\hat{b}_t\right)+\text{H.c}.\right)-2\sqrt{\gamma_2}\left(\hat{\sigma}_z^{(1)}\hat{\sigma}_+^{(2)}\left(\hat{a}_{t-\tau} \text{e}^{\text{i}\theta_1}+\hat{b}_{t+\tau} \text{e}^{-\text{i}\theta_1}\right)\text{e}^{\text{i}\Delta_{12} t}+\text{H.c}\right),
\end{align}
where we have omitted the time dependence in the operators for clarity. Note that by taking the Hermitian conjugate of Eq. (A.4), one straightforwardly obtains the operator equation for $\hat{\sigma}_+^{(1)}$, and hence is not displayed here, similarly for some of the other operators in the list. In the equations, we introduced the Fourier transform of the operator $\hat{a}_\omega$ as
\begin{equation}
\hat{a}_t=\frac{1}{\sqrt{2\pi}}\int_0^\infty \text{d}\omega \hat{a}_w(0)\text{e}^{-\text{i}(\omega-\omega_1)t},
\end{equation}
and similarly for $\hat{b}_t$. We study a coherent input pulse incident from left
\begin{equation}
|\psi\rangle=|\alpha,0_b,\text{g}_1,\text{g}_2\rangle=\text{exp}\left(\eta \hat{A}^\dagger-\eta^* \hat{A}\right)\ket{\text{vac}},
\end{equation}
where $\ket{\text{vac}}=\ket{0_a}\ket{0_b}\ket{\text{g}_1}\ket{\text{g}_2}$ is the vacuum state of both atoms being in the ground state and no photon. The mean photon number in this pulse is $|\eta|^2$. The forward-propagating photon mode creation operator is defined as
\begin{equation}
\hat{A}^\dagger=\int\text{d}t\xi_a(t)\hat{a}_t^\dagger=\int\text{d}\omega f_a(\omega)\hat{a}^\dagger_\omega,
\end{equation}
and similarly for the backward-propagating photon mode. Here $\xi(t)$ is the temporal shape of the wave packet and $f(\omega)$ is the spectral distribution function, related to each other by the Fourier transform
\begin{equation}
\xi(t)=\frac{1}{\sqrt{2\pi}}\int \text{d}\omega f(\omega)\text{e}^{-\text{i}(\omega-\omega_0)t}.
\end{equation}
We find the action of $\hat{a}_t$ as
\begin{equation}
\hat{a}_t|\alpha\rangle=\eta \text{e}^{-\text{i}\Delta_1 t}\xi(t)|\alpha\rangle.
\end{equation}
Similarly, one has
\begin{equation}
\hat{a}_{t-\tau}|\alpha\rangle=\eta \text{e}^{-\text{i}\Delta_1(t-\tau) }\xi(t-\tau)|\alpha\rangle.
\end{equation}
In the following, we consider a square pulse
\begin{equation}
\xi(t) =
  \begin{cases}
   \sqrt{\frac{\Omega}{2}}       & \text{for} \;\;0\leq t \leq \frac{2}{\Omega}, \\
    0  &  \text{otherwise}.\\
  \end{cases}
\end{equation}
The mean photon flux is then given by
\begin{equation}
|\alpha|^2=\frac{|\eta|^2}{2/\Omega}.
\end{equation}
Now working with the expectation values of the operators, we have
\begin{equation}
\frac{\text{d}}{\text{d}t}\left\langle\sigma_-^{(1)}\right\rangle=-\gamma_1 \left\langle\sigma_-^{(1)}\right\rangle+\sqrt{\gamma_1\gamma_2} \text{e}^{-\text{i}(\Delta_1-\Delta_2)t}\text{e}^{\text{i}\theta_2}\left\langle\sigma_z^{(1)}\sigma_-^{(2)}\right\rangle+\sqrt{\gamma_1}\eta \text{e}^{-\text{i}\Delta_1 t}\xi(t)\left\langle\sigma_z^{(1)}\right\rangle,
\end{equation}
and eight more differential equations.  Solving this set of closed equations with \textit{Mathematica} allows us to obtain the transmittance
\begin{equation}
T=\lim_{t\to\infty}\left(1-\frac{N_\text{ref}(t)}{|\eta|^2}\right),
\end{equation}
where the number of reflected photon, $N_\text{ref}$, is given by
\begin{align}
N_\text{ref}(t)&=\int_0^\infty \text{d} \omega \langle\psi|\hat{b}_\omega^\dagger(t)\hat{b}_\omega(t)|\psi\rangle\\
&=\frac{1}{2}\int_0^t \text{d} t'\left[\gamma_1 \left(\left\langle\sigma_z^{(1)}(t')\right\rangle+1\right)+\gamma_2 \left(\left\langle\sigma_z^{(2)}(t')\right\rangle+1\right)+2\sqrt{\gamma_1\gamma_2}\left(\left\langle\sigma_+^{(1)}\sigma_-^{(2)}(t')\right\rangle\text{e}^{-\text{i}\Delta_{12}t'}\text{e}^{\text{i}\theta_2}+\text{c.c}.\right) \right].
\end{align} 
In the monochromatic regime, the length of the pulse is very long so it suffices to consider the steady state solution. We can then obtain the fraction of light transmitted when the light is injected from the left as given by Eq.~(2) in the main text.
In addition to the results presented in the main text, we also considered the case with arbitrary $\gamma_1$ and $\gamma_2$. We found that  $\gamma_1=\gamma_2$ is the optimal regime for achieving maximum rectification. 

\section{Single photon pulse}
We give here a detailed derivation of the reflectance in the single photon regime, that is Eq.~(\ref{T_formula}), as given in the main text. We consider a single photon pulse incident from the left, that is
\begin{equation}
\ket{\psi(t=0)}=\ket{1_a}\ket{0_b}\ket{\text{g}_1}\ket{\text{g}_2}=\hat{A}^\dagger\ket{\text{vac}}.
\end{equation}

In the single-excitation domain, any state of the system can be decomposed as
\begin{equation}
\ket{\psi(t)}=\int_0^\infty \text{d} \omega c_a(\omega,t)\hat{a}_\omega^\dagger\ket{\text{vac}}+\int_0^\infty \text{d} \omega c_b(\omega,t)\hat{b}_\omega^\dagger\ket{\text{vac}}+c_1(t)\hat{\sigma}_+^{(1)}\ket{\text{vac}}+c_2(t)\hat{\sigma}_+^{(2)}\ket{\text{vac}}.
\end{equation}

We work in a reference frame shifted with respect to the free Hamiltonian, where the Schr\"{o}dinger equations read
\begin{equation}
\dot{c}_a=g_\omega c_1\text{e}^{-\text{i}(\omega_{1}-\omega)t}+g_\omega c_2\text{e}^{-\text{i}(\omega_{2}-\omega)t}\text{e}^{-\text{i}\omega d/v_\text{g}}.
\end{equation}
\begin{equation}
\dot{c}_b=g_\omega c_1\text{e}^{-\text{i}(\omega_{1}-\omega)t}+g_\omega c_2\text{e}^{-\text{i}(\omega_{2}-\omega)t}\text{e}^{\text{i}\omega d/v_\text{g}}.
\end{equation}
\begin{equation}
\dot{c}_1=-\int_0^\infty \text{d}\omega g_\omega(c_a+c_b)\text{e}^{\text{i}(\omega_{1}-\omega)t}.
\end{equation}
\begin{equation}
\dot{c}_2=-\int_0^\infty \text{d}\omega g_\omega \left(c_a \text{e}^{\text{i}\omega d/v_\text{g}}+c_b \text{e}^{-\text{i}\omega d/v_\text{g}}\right)\text{e}^{\text{i}(\omega_{2}-\omega)t}.
\end{equation}
Here without loss of any generality, we have put $x_1=0$. We also omitted the time dependence of the field and atoms variables for clarity. Formally integrating the field variables and substituting them into the equations for the atom variables, one gets a closed set of equations for the atomic variables under the Weisskopf-Wigner approximation
\begin{equation}
\dot{c}_1(t)=-\gamma\left(c_1(t)+c_2(t)\text{e}^{-\text{i}(\Delta_1-\Delta_2) t}\text{e}^{\text{i}\omega_{2}d/v_\text{g}}\right)-\sqrt{\gamma} \text{e}^{-\text{i}\Delta_1 t}\xi(t), \label{1PhotonC1}
\end{equation} 
\begin{equation}
\dot{c}_2(t)=-\gamma\left(c_2(t)+c_1(t)\text{e}^{\text{i}(\Delta_1-\Delta_2) t}\text{e}^{\text{i}\omega_{1}d/v_\text{g}}\right)-\sqrt{\gamma} \text{e}^{\text{i}\omega_0 d/v_\text{g}}\text{e}^{-\text{i}\Delta_2 t}\xi(t).\label{1PhotonC2}
\end{equation}
Note that in deriving this set of equations, we have also made the Markovian approximation and neglected the time delay $d/v_\text{g}$ induced by the distance between the two atoms. In the following, we consider a square pulse
\begin{equation}
\xi(t) =
  \begin{cases}
   \sqrt{\frac{\Omega}{2}}       & \text{for} \;\;0\leq t \leq \frac{2}{\Omega} .\\
    0  &  \text{otherwise}.\\
  \end{cases}
\end{equation}
Finally, solving the set of Eqs.~(\ref{1PhotonC1}) and (\ref{1PhotonC2}) allows us to obtain the reflectivity
  \begin{equation}
  R=\lim_{t\rightarrow \infty}N_\text{ref}(t),
  \end{equation}
  with
  \begin{align}
  N_\text{ref}(t)=&\int_0^\infty \text{d}\omega\bra{\psi(t)}\hat{b}_\omega^\dagger \hat{b}_\omega\ket{\psi(t)}\\
  =&\int_0^\infty \text{d} \omega|c_b(\omega)|^2\\
  =&\gamma\int_0^t \text{d} t'\left|c_1(t')+c_2(t')\text{e}^{-\text{i}(\Delta_1-\Delta_2) t'}\text{e}^{\text{i}\omega_{2}d/v_\text{g}} \right|^2.
  \end{align}
  
Focusing on the near-resonant case for both atoms ($\Delta_1\ll\omega_0$ and $\Delta_2\ll\omega_0$), we find that the reflectivity in the monochromatic limit ($\Omega\ll\gamma$) has the form of Eq.~(\ref{T_formula}).

\end{widetext}

\end{document}